\documentclass{kluwer}    
\input{psfig}
\newdisplay{guess}{Conjecture}
\begin{document}                                                                                   
\begin{article}
\begin{opening}         
\title{The formation and evolution of field massive galaxies} 
\author{Andrea \surname{Cimatti}}  
\runningauthor{Andrea Cimatti}
\runningtitle{Evolution of massive galaxies}
\institute{INAF - Osservatorio Astrofisico di Arcetri, Italy}

\begin{abstract}
The problem of the formation and evolution of field massive galaxies is 
briefly reviewed from an observational perspective. The motivations and
the characteristics of the K20 survey are outlined. The redshift 
distribution of $K_s<20$ galaxies, the evolution of the rest-frame 
$K_s$-band luminosity function and luminosity density to $z\sim 1.5$, the 
nature and the role of the red galaxy population are presented. 
Such results are compared with the predictions of models of galaxy evolution.
\end{abstract}

\end{opening}           

\section{Introduction}  
Despite the recent developments in observational cosmology, one of the 
main unsolved issues remains how and when the present-day massive
elliptical
galaxies (${\cal M}_{\rm stars}>10^{11}$ M$_{\odot}$) built up and what 
type of evolution characterized their growth across the cosmic time.

There are two main proposed scenarios. In the first, such systems formed
at high redshifts (e.g. $z>2 - 3$) through a ``monolithic'' collapse 
accompanied by a violent burst of star formation, then followed by a 
passive and pure luminosity evolution (PLE) of the stellar population to 
nowadays (Eggen, Lynden-Bell \& Sandage 1962; Tinsley 1972; Larson 1974; 
van Albada 1982). Such a scenario makes some critical and rigid predictions 
that can
be tested with the observations: {\it (i)} the comoving number density of 
massive spheroids should be conserved through cosmic times, {\it (ii)} 
massive galaxies should evolve only in luminosity, {\it (iii)} old 
passively evolving spheroids should exist at least up to $z\sim1-1.5$, 
{\it (iv)} there should be a population of progenitors at $z>2 - 3$ 
characterized by large amounts of gas (and dust) and strong star formation
rates in order to be compatible with the rapid formation scenario and 
with the properties (e.g. masses, ages, metallicities) of the 
present-day ``fossils'' resulting from that formation process.

In a diametrically opposed scenario, massive spheroids formed at later
times through a slower process of hierarchical merging of smaller galaxies 
(e.g. White \& Rees 1978; Kauffman, White, \& Guiderdoni 1993; Kauffmann
1996) characterized 
by moderate star formation rates, thus reaching the final masses in more recent 
epochs (e.g. $z< 1 - 1.5$) (e.g. Baugh et al. 1996, 1998; Cole et al. 
2000; Baugh et al. 2002). As a consequence, the hierarchical merging models 
(HMMs) predict that massive systems should be very rare at $z\sim1$, with the 
comoving density 
of ${\cal M}_{\rm stars}>10^{11}$ M$_{\odot}$ galaxies decreasing by almost an 
order of magnitude from $z\sim0$ to $z\sim1$ (Baugh et al. 2002; 
Benson et al. 2002). 

Several observations were designed over the recent years in order to
test such two competing models. 

One possibility is to search for the starburst progenitors expected
at $z> 2 - 3$ in the ``monolithic''+PLE scenario. In this respect, 
submm and mm
continuum surveys unveiled a population of high-$z$ dusty starbursts
which may represent the ancestors of the present-day massive galaxies
(see Blain et al. 2002 for a recent review).

The other possibility is to search for passively evolving spheroids 
to the highest possible redshifts and to study their properties both 
in clusters and in the field. This latter approach provided so far 
controversial results.

Because of their color evolution, fundamental plane and stellar population 
properties, cluster ellipticals are now generally believed to form a 
homogeneous population of old systems formed at high redshifts (e.g. 
Stanford et al. 1998; see also Renzini 1999; Renzini \& Cimatti 1999; 
Peebles 2002 for recent reviews). 

However, the question of field spheroids is still actively debated. It is 
now established that old, passive and massive systems exist in the field 
out to $z\sim 1.5$ (e.g. Spinrad et al. 1997; Stiavelli et al. 1999; Waddington 
et al. 2002), but the 
open question is what are their number density and physical/evolutionary 
properties with respect to the model predictions. 

Some surveys based on color or morphological selections found 
a deficit of $z>1 - 1.4$ elliptical candidates (e.g. Kauffmann et al.
1996; Zepf 1997; 
Franceschini et al. 1998; Barger et al. 1999; Rodighiero et al. 2001;
Smith et al. 2002; Roche et al. 2002), whereas others did not confirm 
such result out to $z\sim 1-2$ (e.g. Totani \& Yoshii 1998;
Benitez et al. 1999; Daddi et al. 
2000b; Im et al. 2002; Cimatti et al. 2002a). Part of the discrepancies 
can be ascribed to the strong clustering (hence field-to-field
variations) of the galaxies with the red colors expected for 
high-$z$ elliptical candidates (Daddi et al. 2000a). 

Other approaches made the picture even more controversial. For
instance, Menanteau et al. (2001) found that a fraction of morphologically 
selected field spheroidals show internal color variations incompatible 
with a traditional PLE scenario and stronger than cluster spheroidals 
at the same redshifts. Similar results have been obtained with photometric,
spectroscopic and fundamental plane studies of field ellipticals to $z\sim
0.7 - 1$ 
(e.g. Kodama et al. 1999; Schade et al. 1999; Treu et al. 2002). Such 
observations suggest that, despite the mass of massive spheroids seems
not to change significantly from $z\sim1$ to $z\sim0$ (Brinchmann \&
Ellis 2000), field early-type systems at $z \sim 0.5 - 1$ do 
not form an entirely homogeneous population, some looking consistent with 
the PLE scenario, whereas others with signatures of recent secondary 
episodes of star formation (see also Ellis 2000 for a review). 

A more solid and unbiased approach is to investigate the evolution 
of massive galaxies by means of spectroscopic surveys of field galaxies 
selected in the $K$-band (e.g. Broadhurst et al. 1992), and to
push the study of massive systems to $z>1$. Since the 
rest-frame optical and near-IR light is a good tracer of the galaxy 
{\it stellar} mass (Gavazzi et al. 1996), $K$-band surveys provide the 
important possibility to select galaxies according to their mass up to 
$z\sim2$. The advantages of the $K$-band selection also include the small
k-corrections with respect to optical surveys (which are sensitive to
the star formation activity rather than to the stellar mass), and the
minor effects of dust extinction. Once a sample of faint field galaxies
has been selected in the $K$-band, deep spectroscopy with 8-10m class 
telescopes can then be performed to shed light on their nature and on their 
redshift distribution. Several spectroscopic surveys of this kind have 
been and are being performed (e.g. Cowie et al. 1996; Cohen et 
al. 1999; Stern et al. 2001; see also Drory et al. 2001, although mostly
based on photometric redshifts). 

In this paper, the main results obtained so far with a new spectroscopic 
survey for $K$-selected field galaxies are reviewed, concentrating on 
the redshift distribution, the evolution of 
the near-IR luminosity function and luminosity density, the very red
galaxy population, and on the comparison with the predictions of the
most recent scenarios of galaxy formation and evolution.
$H_0=70$ km s$^{-1}$ Mpc$^{-1}$, $\Omega_{\rm m}=0.3$ and 
$\Omega_{\Lambda}=0.7$ are adopted. 

\section{The K20 survey}

Motivated by the above open questions, we started an ESO VLT Large 
Program (dubbed ``K20 survey'') based on 17 nights distributed over two 
years (1999-2000) (see Cimatti et al. 2002c for details). 

The prime aim of such a survey was to derive the redshift distribution and
spectral properties of 546 $K_s$-selected objects with the {\it only}
selection criterion of $K_s<20$ (Vega). Such a threshold is 
critical because it selects galaxies over a broad range 
of masses, i.e. ${\cal M}_{\rm stars}>10^{10}$ M$_{\odot}$ 
and ${\cal M}_{\rm stars}>4 \times10^{10}$ M$_{\odot}$ for $z=0.5$ and 
$z=1$ respectively (according to the mean ${\cal M}_{\rm stars}/L$ 
ratio in the local universe and adopting Bruzual \& Charlot 2000 spectral 
synthesis models with a Salpeter IMF). The $K_s<20$ selection has also the 
observational advantage that most galaxies have magnitudes still within the 
limits of optical spectroscopy of 8m-class telescopes ($R<25$).

The targets were selected from $K_s$-band images (ESO NTT+SOFI) of
{\it two independent fields} covering a total area of 52 arcmin$^2$.
One of the fields is a
sub-area of the Chandra Deep Field South (CDFS; Giacconi et al. 2001). 
Optical multi-object spectroscopy was made with the ESO VLT 
UT1 and UT2 equipped with FORS1 and FORS2. A fraction of the sample was 
also observed with near-IR spectroscopy with VLT UT1+ISAAC in order to 
attempt to derive the redshifts of the galaxies which were too faint 
for optical spectroscopy and/or expected to be in a redshift range
for which no strong features fall in the observed optical spectral
region (e.g. $1.5<z<2.0$). In addition to spectroscopy, $UBVRIzJK_s$ imaging 
was also available for both fields, thus providing the possibility to
estimate photometric redshifts for all the objects in the K20 sample, to
optimize them through a comparison with the spectroscopic redshifts and 
to assign a reliable photometric redshift to the objects for which it was not
possible to derive the spectroscopic z. The overall spectroscopic redshift 
completeness is 94\%, 92\%, 87\% for $K_s<19.0$, 19.5, 20.0 respectively. 
The overall redshift completeness (spectroscopic + photometric redshifts) 
is 98\%.

The K20 survey represents a significant improvement with respect to
previous surveys for faint $K$-selected galaxies (e.g. Cowie et al. 1996;
Cohen et al. 1999) thanks to its larger sample, the coverage of two
independent fields (thus reducing the cosmic variance effects), the
availability of optimized photometric redshifts, and the spectroscopic 
redshift completeness, in particular for the reddest galaxies. 

\section{The redshift distribution of $K_s<20$ galaxies}

The observed differential and cumulative redshift distributions for the 
K20 sample are presented in Fig. 1 (see Cimatti et al.
2002b), together with the predictions of different scenarios of galaxy 
formation and evolution, including both hierarchical merging models 
(HMMs) from Menci et al. (2002, M02), Cole et al. (2000, C00), 
Somerville et al. (2001, S01), and pure luminosity evolution models 
(PLE) based on Pozzetti et al.  (1996,1998 PPLE) and Totani et al. (2001, TPLE).
The redshift distribution can be retrieved from {\it http://www.arcetri.
astro.it/$\sim$k20/releases}. The spike at $z\sim 0.7$ is due to two clusters 
(or rich groups) at $z=0.67$ and $z=0.73$. The median redshift of $N(z)$
is $z_{med}=0.737$ and $z_{med}=0.805$, respectively with and without
the two clusters being included. Without the clusters, the fractions
of galaxies at $z>1$ and $z>1.5$ are 138/424 (32.5\%) and 39/424
(9.2\%) respectively. The high-$z$ tail extends beyond $z=2$.
The contribution of objects with only a photometric redshift
becomes relevant only for $z>1.5$. The fractional cumulative distributions 
displayed in Fig. 1 (bottom panels) were obtained by 
removing the two clusters mentioned above in order to perform a meaningful 
comparison with the galaxy formation models which do not include clusters 
(PLE models), or are averaged over very large volumes, hence diluting the 
effects of redshift spikes (HMMs). No best tuning of the models was 
attempted in this comparison, thus allowing an unbiased {\it blind test}
with the K20 observational data. The model predicted $N(z)$ are 
normalized to the K20 survey sky area. 

\begin{figure} 
\begin{center}
\begin{tabular}{c}
\psfig{figure=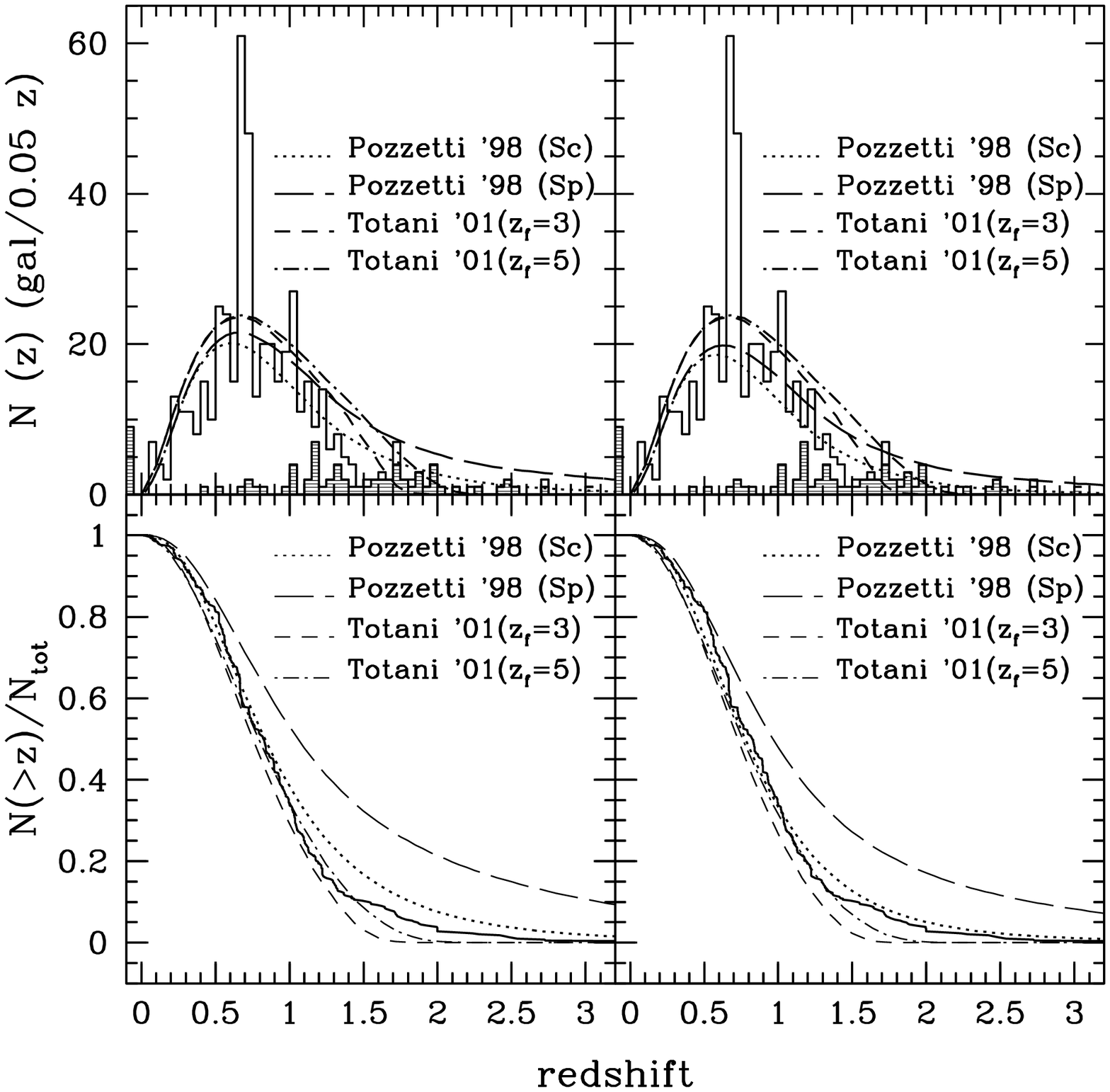,height=6cm}
\psfig{figure=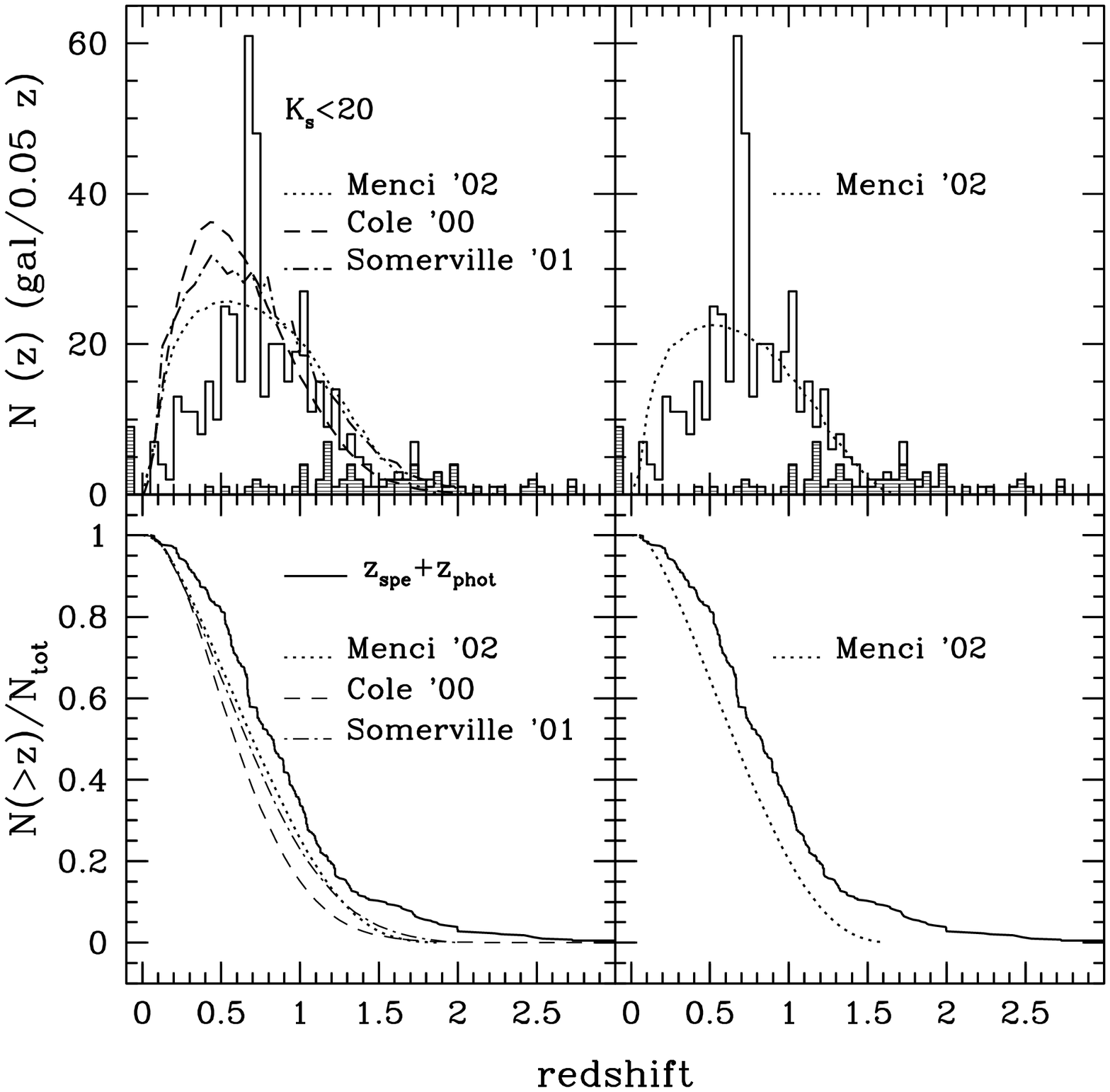,height=6cm}
\end{tabular}
\end{center}
\caption[]{{\bf Fig. 1a --} {\it Top panels:} the observed differential 
$N(z)$ for
$K_s<20$ (histogram) compared with the PLE model predictions.
{\it Bottom panels:} the observed fractional cumulative redshift
distribution (continuous line) compared with the same models.
The shaded histogram shows the contribution of photometric redshifts.
The bin at $z<0$ indicates the 9 objects without redshift.
The {\it left} and {\it right} panels show the models without and with
the inclusion of the photometric selection effects respectively. 
Sc and Sp indicate Scalo and Salpeter IMFs respectively.

{\bf Fig. 1b --} same as Fig. 1a, but compared with the HMM predictions. 
{\it Right} panels: the M02 model with the inclusion of the photometric 
selection effects.
}
\end{figure}

\begin{figure} 
\begin{center}
\begin{tabular}{c}
\psfig{figure=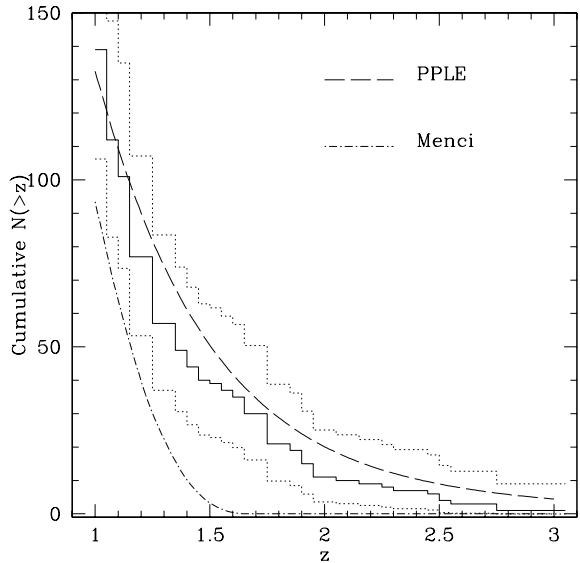,height=8cm}
\end{tabular}
\end{center}
\caption[]{
The observed cumulative {\it number} of galaxies between $1<z<3$
(continuous line) and the corresponding poissonian $\pm 3 \sigma$
confidence region (dotted lines). The PPLE (Scalo IMF)
and the M02 models are corrected for the photometric biases.
}
\end{figure}

Fig. 1a shows a fairly good agreement between the observed
$N(z)$ distribution and the PLE models (with the exception of PPLE
with Salpeter IMF), although such models slightly overpredict the number 
of galaxies at $z\gsim 1.2$. However, if the photometric selection
effects present in the K20 survey (Cimatti et al. 2002b) are taken 
into account, the PLE models become much closer to the observed $N(z)$ 
thanks to the decrease of the predicted high-$z$ tail. According to the 
Kolmogorov-Smirnov test, the PLE models are acceptable at 95\% confidence 
level, with the exception of the PPLE model with Salpeter IMF. 

On the other side, all the HMMs underpredict
the median redshift ($z_{med}$=0.59, 0.70 and 0.67 for the C00, M02
and S01 models respectively), overpredict the total number of galaxies 
with $K_s<20$ by factors up to $\sim$50\% as well as the number of galaxies at
$z<0.5$, and underpredict the fractions of $z>1 - 1.5$ galaxies 
by factors of $2 - 4$ (Fig. 1b). Fig. 1b (bottom panels) 
illustrates that in the fractional 
cumulative distributions the discrepancy with observations appears
systematic at all redshifts. The Kolmogorov-Smirnov test shows
that all the HMMs are discrepant with the observations at $>99$\%
level. The inclusion of the photometric biases exacerbates this
discrepancy, as shown in Fig. 1b (right panels) for the M02 model
(the discrepancy for the C00 and S01 models becomes even stronger). 
The deficit of high-redshift objects is well
illustrated by Fig. 2, where the PPLE 
model is capable to reproduce the cumulative {\it number} distribution of 
galaxies at $1<z<3$ within 1-2$\sigma$, whereas the M02 model is always 
discrepant at $\geq 3\sigma$ level (up to $>5 \sigma$ for $1.5<z<2.5$).
This conclusion is not heavily based on the objects with only 
photometric redshifts estimates, as the mere presence of 7 galaxies 
with spectroscopic redshift $z>1.6$ is already in substantial contrast
with the predictions by HMMs of basically no galaxies with $K_s<20$ 
and $z>1.6$.

\section{The evolution of the luminosity function} 

The luminosity function (LF) of galaxies has been estimated in the rest-frame
$K_s$-band and in three redshift bins which avoid the clusters at
$z\sim0.7$ ($z_{mean}$=0.5,1,1.5; see Fig. 3) (Pozzetti et al. 
2003), using both the 1/$V_{max}$ (Schmidt 1968; Felten 1976) 
and the STY (Sandage, Tammann \& Yahil 1979) formalisms. 
The LF observed in the first two redshift bins is fairly well fit by 
Schechter functions. A comparison with the local $K_s$-band LF of Cole 
et al. (2001) shows a mild {\it luminosity} evolution of LF($z$) 
out to $z=1$, with a brightening of about -0.5 magnitudes from $z=0$
to $z=1$ (Fig. 3). Similar results have been found by Drory et al.
(2001), Cohen (2002), Bolzonella et al. (2002) and Miyazaki et al. 
(2002) (see also Cowie et al. 1996).

The study of the LF by galaxy spectral or color types shows that
red early-type galaxies dominate the bright-end of the LF already
at $z\sim1$, and that their number density shows only a small decrease
from $z\sim0$ to $z\sim1$ (Pozzetti et al. 2003). This is consistent
with the independent study of Im et al. (2002) based on morphologically
selected spheroidals.

\begin{figure} 
\begin{center}
\begin{tabular}{c}
\psfig{figure=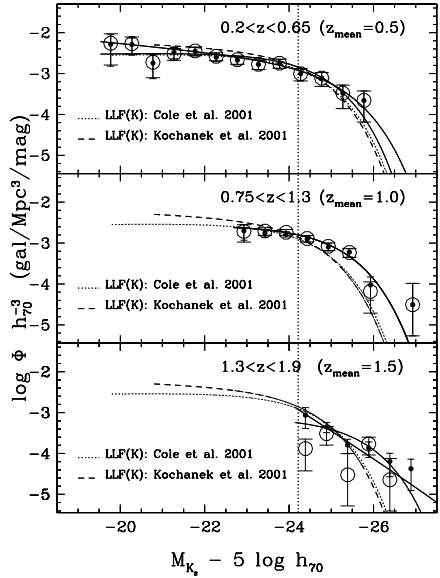,height=8cm}
\end{tabular}
\end{center}
\caption[]{
The rest-frame $K_s$-band Luminosity Function in three redshift bins.
Data points derive from $1/V_{max}$ analysis. Solid curves: the Schechter 
fits derived from maximum likelihood analysis (thin solid lines are the 
fit assuming local $\alpha$ parameter). Dotted and dashed curves: the 
local $K_s$-band LFs of Cole et al. (2001) and Kochanek et al. (2001)
respectively. The vertical dotted line indicates the local $M^{\ast}$
of Cole et al. (2000). Open circles: spectroscopic redshifts, filled 
circles: spectroscopic + photometric redshifts.
}
\end{figure}

Fig. 4 shows a comparison of the observed luminosity function with PLE and 
HMM predictions. The PLE models describe reasonably 
well the shape and the evolution of the luminosity function up to the 
highest redshift bin, $z_{mean}=1.5$, with no evidence for a strong decline 
of the most luminous systems (with $L>L^*$). This is in contrast, especially 
in the highest redshift bin, with the prediction by the HMMs of a decline in 
the number density of luminous (i.e. massive) systems with redshift.
Moreover, hierarchical merging models 
(namely M02 and C00) result in a significant overprediction of faint, 
sub--$L^{*}$ galaxies at $0<z<1.3$. This problem, also hinted by the 
comparison of $N(z)$ between models and data, is probably related to 
the so called ``satellite problem" (e.g.Primack 2002). 

However, it is interesting to note that at $z\sim 1$ the HMMs seem not
to be in strong disagreement with the observations relative to the bright 
end of the galaxy luminosity function (with the possible exception of the
highest luminosity point). Thus, the key issue is to verify 
whether the bright $L>L^*$ galaxies in the K20 survey have the same nature 
of the luminous galaxies predicted by the HMMs, in particular for their 
mass to light ratios (${\cal M}_{\rm stars}/L$).

\begin{figure} 
\begin{center}
\begin{tabular}{c}
\psfig{figure=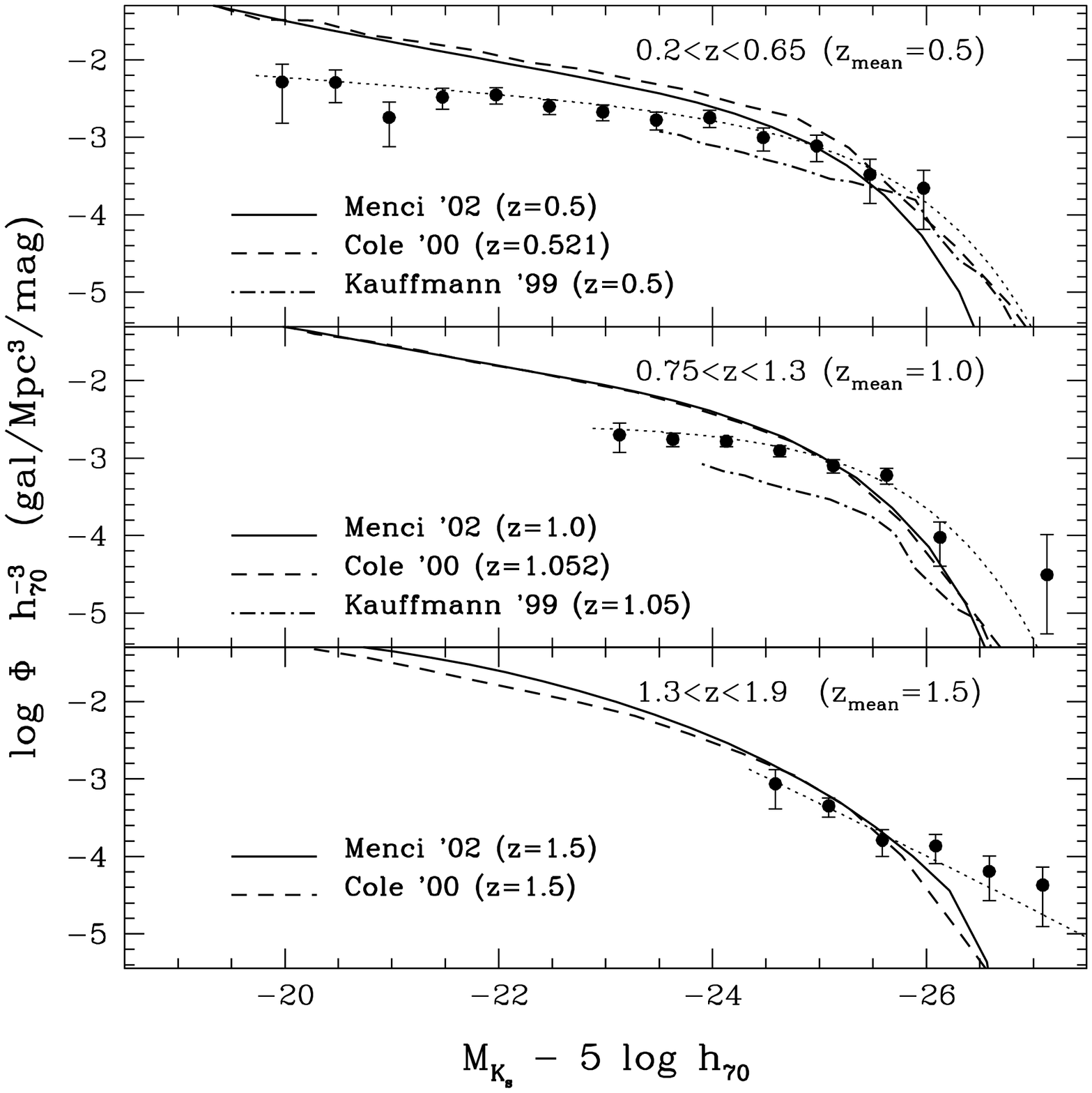,height=6cm}
\psfig{figure=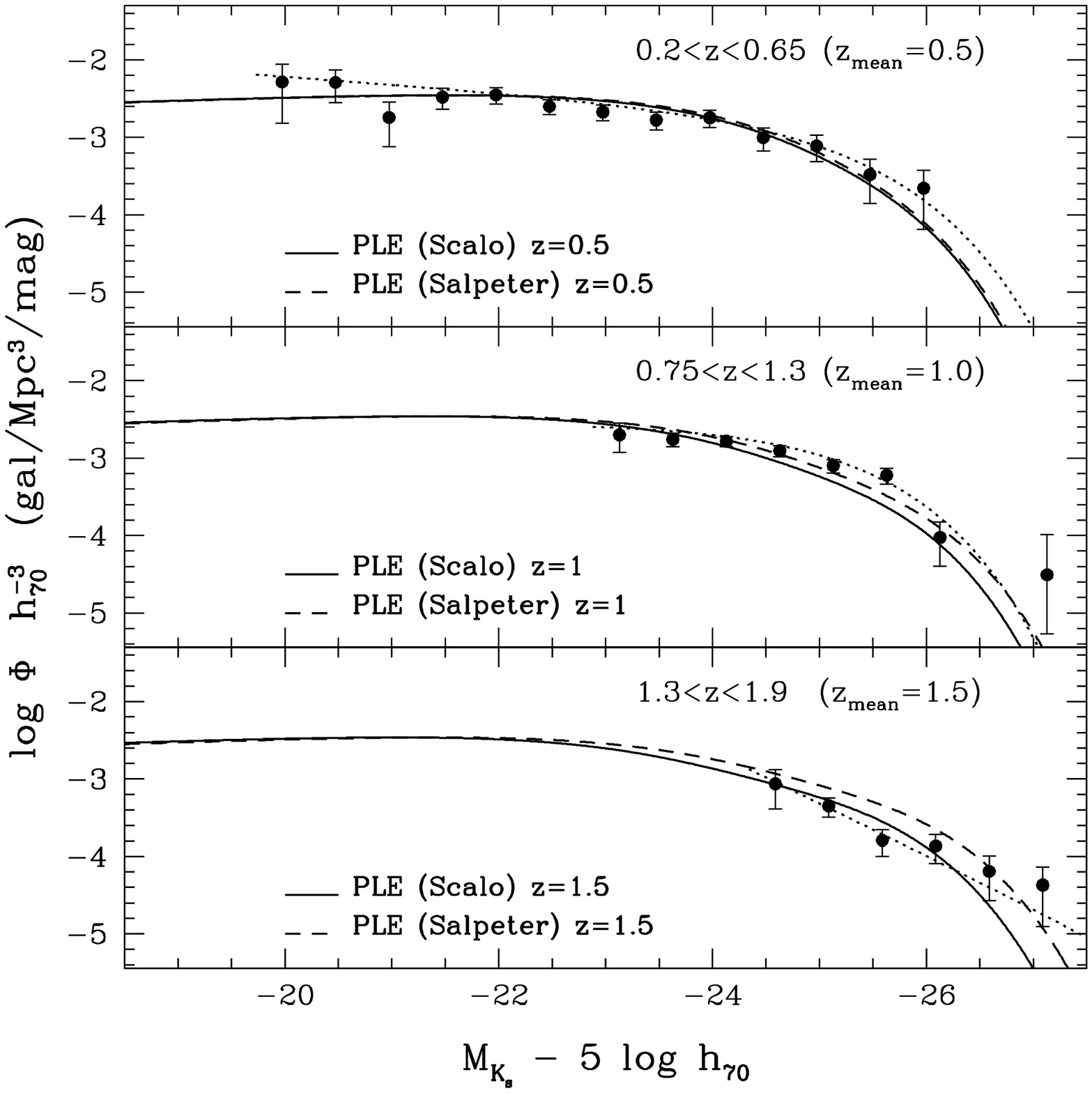,height=6cm}
\end{tabular}
\end{center}
\caption[]{Left: the $K_s$-band LF compared to hierarchical merging model
predictions. Right: the $K_s$-band LF compared to PLE (PPLE) model 
predictions. Dotted curves are the Schechter best fits to the observed
LFs (spectroscopic + photometric redshifts).
}
\end{figure}

\begin{figure} 
\begin{center}
\begin{tabular}{c}
\psfig{figure=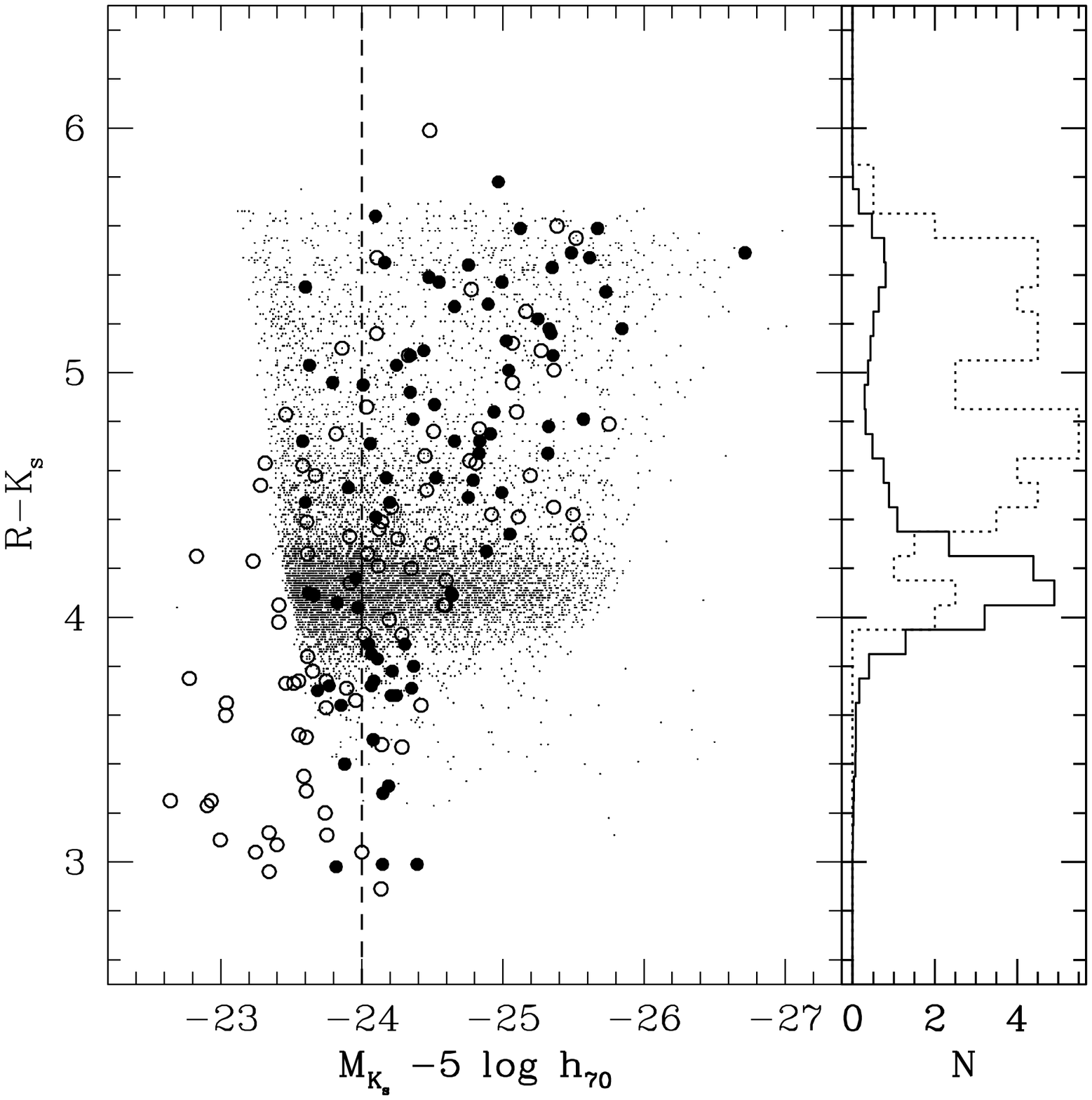,height=8cm}
\end{tabular}
\end{center}
\caption[]{
{\it Left panel:} $R-K_s$ colors vs. rest-frame absolute $K_s$
magnitudes for $z=1.05$ GIF simulated catalog (small dots)
and data (circles) at $0.75<z<1.3$ (spectroscopic + photometric
redshifts; $z_{mean}=1$)
(empty and filled circles refer to $z<1$ and $z>1$ respectively).
The vertical dashed line represents approximately the completeness magnitude
limit of GIF catalog corresponding to its mass limit (see text).
{\it Right panel:} Color distribution of luminous galaxies ($M_{K_s}-5$ 
log $h_{70}<-24.5$) observed (dotted line) and simulated (continuous line), 
normalized to the same comoving volume.
}
\end{figure}

Fig. 5 compares the $R-K_s$ colors and luminosity distributions of galaxies 
with $0.75<z<1.3$ (a bin dominated by spectroscopic redshifts) as observed 
in our survey to the predictions of the GIF\footnote{http://
www.mpa-garching.mpg.de/GIF/} simulations (Kauffmann et al. 
1999). Such a comparison highlights that a relevant discrepancy is present 
between the two distributions: real galaxies with $M_K-5$log$h_{70}<-24.5$ in
the K20 sample have a median color of $R-K_s\sim5$, whereas the GIF
simulated galaxies have $R-K_s\sim4$, and the two distributions have very 
small overlap. Given that red galaxies have old stellar populations and 
higher ${\cal M}_{\rm stars}/L$ ratios, the apparent agreement with HMM 
predictions of the $z\sim1$ bright end of the luminosity function (Fig. 4) is 
fortuitous and probably results from an underestimate of the 
${\cal M}_{\rm stars}/L$ present in the same models. This 
is equivalent to say that the number density of massive galaxies at 
$z\sim1$ is underpredicted by HMMs, and the predicted colors, ages and 
star formation rates do not agree with the observations.

\section{The evolution of the luminosity density}

Tracing the integrated cosmic emission history of the galaxies at 
different wavelengths offers the prospect of an empirical determination of 
the global evolution of the galaxy population. Indeed it is independent 
of the details of galaxy evolution and depends mainly on the star formation 
history of the universe (Lilly et al. 1996, Madau, Pozzetti \& Dickinson
1998). Attempts to reconstruct the cosmic evolution of the comoving 
luminosity density have been made previously mainly in the UV and optical 
bands, i.e. focusing on the star formation history activity of galaxies
(Lilly et al. 1996, Cowie et al. 1999). 

Our survey offers for the first time the possibility to investigate it 
in the near-IR using a LF extended over a wide range in luminosity, thus
providing new clues on the global evolution of the stellar mass density
(Pozzetti et al. 2003).
Using the local luminosity density at $z\sim0$ as derived from Cole et al.
(2001) complemented with the estimates at higher redshifts based on the K20
survey, it is found that the rest-frame $K_s$-band luminosity density up 
to $z\sim1.3$ is well represented by a power law with $\rho_\lambda(z)= 
\rho_\lambda(z=0) (1+z)^{\beta}$, with $\beta=0.37$. 
Compared to the optical (rest-frame UV-blue) bands, the near-IR luminosity 
density evolution is much slower ($\beta=3.9-2.7$ from 0.28 to 0.44 
$\mu$m by Lilly et al. 1996 and $\beta=1.5$ at 0.15-0.28 $\mu$m by 
Cowie et al. 1999, for $\Omega_m=1$). 
The slow evolution of the observed $K_s$-band luminosity density suggests
that the stellar mass density should also evolve slowly at least up to 
$z\sim1.3$. This is in agreement with a recent analysis by Bolzonella et 
al. (2002) (see also Cowie et al. 1996 and Brinchmann \& Ellis 2000). 
The analysis of the stellar mass function and its cosmic evolution is 
in progress and will be presented elsewhere.

\section{Extremely Red Objects (EROs)}

Extremely Red Objects (EROs, $R-K>5$) are critical in the context
of galaxy formation and evolution because their colors allow to select
old and passively evolving galaxies at $z>0.9$. 

For a fraction of EROs (70\% to $K_s<19.2$) present in the K20 sample it 
was possible to derive a spectroscopic redshift and a spectral 
classification (Cimatti et al 2002a). Two classes of galaxies
at $z\sim1$ contribute nearly equally to the ERO population: 
old stellar systems with no signs of star formation, and dusty 
star-forming galaxies. 

\begin{figure} 
\begin{center}
\begin{tabular}{c}
\psfig{figure=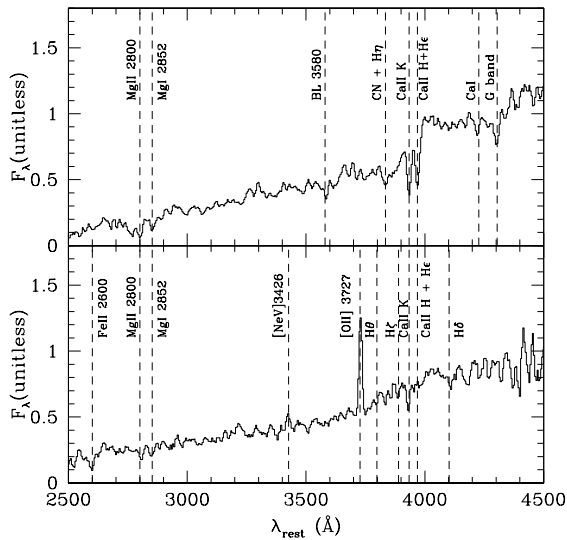,height=8cm}
\end{tabular}
\end{center}
\caption[]{
The average rest-frame spectra (smoothed with a 3 pixel boxcar) of
old passively evolving (top; $z_{mean}=1.000$) and dusty
star-forming EROs (bottom; $z_{mean}=1.096$) with $K_s\leq20$ (Cimatti et
al. 2002a).}
\end{figure}

\subsection{Old EROs}

The colors and spectral properties of old EROs are consistent with 
$\geq$3 Gyr old passively evolving stellar populations (assuming 
solar metallicity and 
Salpeter IMF), requiring a formation redshift $z_f>2.4$. The number
density is $6.3\pm1.8\times10^{-4}$h$^{3}$Mpc$^{-3}$ for $K_s<19.2$, 
consistent with the expectations of PLE models for passively evolving 
early-type galaxies with similar formation redshifts (Cimatti et al. 
2002a). HMMs predict a significant deficit of such old red galaxies at 
$z\sim1$, ranging from a factor of $\sim3$ (Kauffmann et al. 1999) to 
a factor of $\sim5$ (Cole et al. 2000). Preliminary analysis of recent 
HST+ACS imaging shows that old EROs have indeed spheroidal morphologies 
with surface brightness profiles typical of elliptical galaxies. 

\subsection{Dusty star-forming EROs}

The spectra of star-forming EROs suggest a dust reddening of 
$E(B-V)\sim0.5$--1 (adopting the Calzetti 2001 extinction law),
implying typical star-formation rates of 50-150 M$_\odot$yr$^{-1}$, 
and a significant contribution ($>20-30$\%) to the cosmic star-formation 
density at $z\sim1$ (see also Smail et al. 2002). A recent analysis based 
on their X-ray emission provided a similar estimate of the SFRs 
(Brusa et al. 2002). 

The comoving density of dusty EROs is again $\sim6\times10^{-4}$
h$^{3}$Mpc$^{-3}$ at $K_s<19.2$. The GIF simulations (Kauffmann et al. 
1999) predict a comoving density of red galaxies with $SFR>50$ 
M$_\odot$yr$^{-1}$ that is a factor of 30 lower than the observed 
density of dusty EROs.

Such moderate SFRs suggest that the far-infrared luminosities
of dusty star-forming EROs are generally below $L_{FIR}\sim10^{12}$ 
L$_{\odot}$, and would then explain the origin of the low detection 
rate of EROs with $Ks<20 - 20.5$ in submm continuum observations
(e.g. Mohan et al. 2002; see also Smail et al. 2002). However, the 
fraction of dusty ultraluminous infrared systems may be higher in 
ERO samples selected at fainter $Ks$-band magnitudes (e.g. Wehner 
et al. 2002).

\subsection{Clustering}

Taking advantage of the spectroscopic redshift information for the two 
ERO classes, we compared the relative 3D clustering in real space (Daddi 
et al. 2002). The comoving correlation lengths 
of dusty and old EROs are constrained to be $r_0<2.5$ and $5.5<r_0<16 \
h^{-1}$ Mpc comoving respectively, implying that old EROs are the main 
source of the ERO strong angular clustering. It is important to notice that 
the strong clustering measured for the old EROs is in agreement with 
the predictions of hierarchical clustering scenarios (Kauffmann et al. 1999).

\section{Summary and discussion}

The high level of completeness of the K20 survey and the relative set of 
results presented in previous sections provide new implications for 
a better understanding of the evolution of ``mass-selected" field galaxies.

\noindent
{\bf (1)} The redshift distribution of $K_s<20$ field galaxies has a median 
redshift of $z_{med}\sim0.8$ and a high-$z$ tail extended beyond $z\sim2$.
The current models of hierarchical merging do not match the observed median
redshift because they significantly overpredict the number of low luminosity 
(hence low mass) galaxies at $z<0.4 - 0.5$, and 
underpredict the fraction of objects at $z>1 - 1.5$. Instead, the 
redshift distributions predicted by PLE models are in reasonable agreement 
with the observations.  
It is relevant to recall here that early predictions of the expected 
fraction of galaxies at $z>1$ in a $K_s<20$ sample indicated respectively 
$\approx 60\%$ and $\approx 10\%$ for a PLE case and for a (then) standard 
$\Omega_m=1$ CDM model (Kauffmann \& Charlot 1998). This version of PLE was
then ruled out by Fontana et al. (1999). The more recent PLE models and 
HMMs consistently show that for $z>1$ the difference 
between the predictions of different scenarios is much less extreme. These 
results come partly from the now favored $\Lambda$CDM cosmology which 
pushes most of the merging activity in hierarchical models at earlier times 
compared to $\tau$CDM and SCDM models with $\Omega_m=1$ (structures form 
later in a matter-dominated universe, thus resulting in an even 
lower fraction of galaxies at high-$z$), and partly to different recipes 
for merging and star formation modes, which tend to narrow the gap between 
HMMs and the PLE case (e.g. Somerville et al.  2001; Menci et al. 2002). 
In this respect, the observed $N(z)$ provides an additional evidence that the 
universe is not matter-dominated ($\Omega_m<1$). 

\noindent
{\bf (2)} The rest-frame $K_s$-band luminosity function shows a mild luminosity
evolution up to at least $z\sim 1$, with a brightening of about 0.5 
magnitudes. Significant density evolution is ruled out up to $z\sim 1$.
Current hierarchical merging models fail in reproducing the shape
and evolutionary properties of the LF because they overpredict the number 
of sub-$L^{\ast}$ galaxies and predict a substantial density evolution.
PLE models are in good agreement with the observations up to $z\sim1$.

\noindent
{\bf (3)} At odds with the HMMs, the bright-end of the LF at $z\sim1$ 
is dominated by red and luminous (hence old and massive) galaxies. 

\noindent
{\bf (4)} The rest-frame $K_s$-band luminosity density (hence the
stellar mass density) evolves slowly up to $z\sim1.3$. 

\noindent
{\bf (5)} Old passive systems and dusty star-forming galaxies
(both at $z\sim1$) equally contribute to the ERO population with
$K_s<19.2$. 

\noindent
{\bf (6)} The number, luminosities and ages of old EROs 
imply that massive spheroids formed at $z>2.4$ and that were
already fully assembled at $z\sim1$, consistently with a PLE scenario. 

\noindent
{\bf (7)} Dusty EROs allow to select (in a way complementary to other
surveys for star-forming systems) a population of galaxies which contribute 
significantly to the cosmic star formation budget at $z\sim1$. 

\noindent
{\bf (8)} HMMs strongly underpredict the number of both ERO classes.

Overall, the results of the K20 survey show that galaxies selected in the
$K_s$-band are characterized by little evolution up to $z\sim1$, and that
the observed properties can be successfully described by a PLE scenario.
In contrast, HMMs fail in reproducing the observations because they
predict a sort of ``delayed'' scenario where the assembly of massive
galaxies occurs later than what is actually observed. We recall
here that the discrepancies of HMMs in accounting for the properties of 
even $z=0\rightarrow\sim 1$ early-type galaxies have been already emphasized 
in the past (e.g., Renzini 1999; Renzini \& Cimatti 1999). Moreover, among 
low-redshift galaxies there appears to be a clear anti-correlation of the 
specific star formation rate with galactic mass (Gavazzi et al. 1996; 
Boselli et al. 2001), the most massive galaxies being ``old'', the low-mass 
galaxies being instead dominated by young stellar populations. This is just 
the opposite than expected in the traditional HMMs, where the most massive 
galaxies are the last to form. The same anti-correlation is observed in the
K20 survey at $z\sim1$. 

{\it It is important to stress here that the above results do not necessarily
mean that the whole framework of hierarchical merging of CDM halos
is under discussion.}
For instance, the strong clustering of old EROs and the clustering
evolution of the K20 galaxies (irrespective of colors) seem to be 
fully consistent with the predictions of CDM models of large scale 
structure evolution (Daddi et al. 2001; Firth et al. 2002; Daddi et al.
in preparation). 

It is also important to stress that the K20 survey allows to perform 
tests which are sensitive to the evolutionary ``modes'' of galaxies rather
than to their formation mechanism. This means that merging, as the galaxy
main formation mechanism, is not ruled out by the present observations. 
Also, it should be noted that PLE models are not a physical alternative 
to the HMMs, but rather tools useful to parameterize the evolution of 
galaxies under three main 
assumptions: high formation redshift, conservation of number density 
through cosmic times, passive luminosity evolution of the stellar 
populations.

Thus, if we still accept the $\Lambda$CDM scenario of hierarchical merging 
of dark 
matter halos as the {\it basic framework for structure and galaxy formation}, 
the observed discrepancies highlighted by the K20 survey may be ascribed 
to how the {\it baryon assembly} is treated and, in particular, to the 
heuristic algorithms adopted for the star formation processes and 
their feedback, both within individual galaxies and in their environment. 
Our results suggest that HMMs should have galaxy formation in a CDM 
dominated universe to closely mimic the old-fashioned {\it monolithic 
collapse} scenario. This requires enhancing merging and star formation in 
massive halos at high redshift (say, $z\gsim 2 - 3$), while in the 
meantime suppressing star formation in low-mass halos. For instance, 
Granato et al. (2001) suggested the strong UV radiation feedback from 
the AGN activity during the era of supermassive black hole formation to 
be responsible for the suppression of star formation in low-mass halos, 
hence imprinting a ``anti-hierarchical'' behavior in the baryonic component. 
The same effect may well result from the feedback by the starburst activity 
itself (see also Ferguson \& Babul 1998).

In summary, the redshift distribution of $K_s<20$ galaxies, together with 
the space density, nature, and clustering properties of the ERO population, 
and the redshift evolution of the rest-frame near-IR luminosity function
and luminosity density provide a new set of observables on the galaxy 
population in the $z\sim 1-2$ universe, thus bridging the properties of 
$z\sim 0$ galaxies with those of Lyman-break and submm/mm-selected galaxies 
at $z\geq 2$--3. This set of observables poses a new challenge for 
theoretical models to properly reproduce.

Deeper spectroscopy coupled with HST+ACS imaging and SIRTF photometry 
will allow us to derive additional constraints on the
nature and evolution of massive stellar systems out to higher redshifts.

\acknowledgements
The K20 survey team includes: S. Cristiani (INAF-Trieste), S. D'Odorico (ESO),
A. Fontana (INAF-Roma), E. Giallongo (INAF-Roma), R. Gilmozzi (ESO), N. Menci 
(INAF-Roma), M. Mignoli (INAF-Bologna), F. Poli (University of Rome), A. 
Renzini (ESO), P. Saracco (INAF-Brera), J. Vernet (INAF-Arcetri), and
G. Zamorani (INAF-Bologna).

We are grateful to C. Baugh, R. Somerville and T. Totani for providing
their model predictions. AC warmly acknowledges Jim Peebles and Mark 
Dickinson for useful and stimulating discussions.

\end{article}
\end{document}